 \documentclass[cameraready]{Interspeech}

\usepackage{textcomp}
\usepackage{amssymb}
\usepackage{CJKutf8}
\usepackage{multirow}
\usepackage{graphicx}
\usepackage{pifont}
\DeclareUnicodeCharacter{FF05}{\%} 
\DeclareUnicodeCharacter{FF06}{\&} 
\DeclareUnicodeCharacter{FF03}{\#} 
\DeclareUnicodeCharacter{FF20}{@}  
\DeclareUnicodeCharacter{FF0A}{*}  
\DeclareUnicodeCharacter{203B}{\textasteriskcentered}
\title{Speech-Worthy Alignment for Japanese SpeechLLMs \\ via Direct Preference Optimization}

\author[]{Mengjie}{Zhao}
\author[]{Lianbo}{Liu}
\author[]{Yusuke}{Fujita}
\author[]{Hao}{Shi}
\author[]{Yuan}{Gao}
\author[]{Roman}{Koshkin}
\author[]{Yui}{Sudo}

\address{
    SB Intuitions, Tokyo, Japan
}

\email{}


\usepackage{comment}

\begin{document}

\maketitle

\begin{abstract}
SpeechLLMs typically combine ASR-trained encoders with text-based LLM backbones, leading them to inherit written-style output patterns unsuitable for text-to-speech synthesis. This mismatch is particularly pronounced in Japanese, where spoken and written registers differ substantially in politeness markers, sentence-final particles, and syntactic complexity. We propose a preference-based alignment approach to adapt Japanese SpeechLLMs for \emph{speech-worthy} outputs: text that is concise, conversational, and readily synthesized as natural speech. To rigorously evaluate this task, we introduce \textbf{SpokenElyza}, a benchmark for Japanese speech-worthiness derived from  ELYZA-tasks-100 with auditory verification by native experts. Experiments show that our approach achieves substantial improvement on SpokenElyza while largely preserving performance on the original written-style evaluation. We will release SpokenElyza to support future research on Japanese spoken dialog systems.
\end{abstract}

\section{Introduction}
Extending large language models (LLMs) with auditory capabilities has enabled numerous real-world applications, including automatic speech recognition (ASR; \cite{radford2023robust,zhang-etal-2023-speechgpt,tang2023salmonn}), sound event understanding \cite{gong2024listen,ghosh-etal-2024-gama}, and music comprehension \cite{ghosh2026music,zhao2024openmu,mao-etal-2025-deepresonance}. These auditory LLMs process audio inputs and generate textual responses, bridging the gap between audio/speech and language understanding.

However, SpeechLLMs like Qwen2-audio \cite{Qwen2-Audio} are typically trained to respond in written text rather than spoken language \cite{cho-etal-2024-speechworthy}. 
This creates a fundamental mismatch for speech-to-speech applications, where the model's output must be synthesized into natural-sounding audio via text-to-speech (TTS) systems. LLMs' written-style outputs often incorporate markdown formatting, bullet points, and complex sentence structures, which are poorly suited for auditory consumption. To address this, \cite{cho-etal-2024-speechworthy} defined \textbf{speech-worthy responses}: text that is concise, avoids non-verbalizable artifacts (e.g., bullet points, markdown), and prioritizes natural conversational flow for auditory comprehension.

Cascaded systems combining SpeechLLMs with TTS remain the dominant paradigm for speech-to-speech interaction. While end-to-end models
such as SLAM-Omni \cite{slamomni} and Llama-Omni \cite{fang2025llamaomni} can generate speech directly, they often suffer from catastrophic forgetting and underperform their cascaded counterparts, e.g., on VoiceBench \cite{chen2024voicebench}. Consequently, producing speech-worthy text from SpeechLLMs remains a critical intermediate step for high-quality spoken dialog systems.
Recent work has explored aligning text-based LLMs to produce speech-worthy outputs  \cite{cho-etal-2024-speechworthy,woo-etal-2025-think}. However, these approaches focus exclusively on text LLMs, leaving the growing ecosystem of SpeechLLMs unexplored.

In this work, we address the challenge of \emph{aligning SpeechLLMs to generate speech-worthy outputs}. SpeechLLMs are typically trained on ASR transcripts or text-based instruction data \cite{chu2024qwen2}, resulting in verbose, structured outputs optimized for reading rather than listening \cite{woo-etal-2025-think}. 
We focus specifically on Japanese which has pronounced stylistic divergence between written and spoken registers.
Natural spoken Japanese requires polite predicates, 
sentence-final particles signaling speaker intent, 
and simpler syntactic structures. 
SpeechLLMs producing grammatically correct written Japanese may nonetheless sound robotic, overly dense, or socially inappropriate when synthesized as speech.
These pronounced divergences make Japanese an ideal testbed for studying speech-worthy alignment.
To the best of our knowledge, no evaluation resources currently exist for Japanese speech-worthy dialog. We address this gap by constructing \textbf{SpokenElyza}, a carefully curated benchmark derived from the widely-used  ELYZA-tasks-100 dataset \cite{elyzatasks100}.

Our contributions are as follows:
\textbf{(1)} We apply preference-based alignment to Japanese SpeechLLMs and demonstrate substantial improvement on speech-worthiness while largely preserving written-style performance. \textbf{(2)} To enable this evaluation, we introduce SpokenElyza, the first benchmark for speech-worthy Japanese spoken dialog, constructed through modality filtering, style transfer, and auditory verification by native experts.
We will release SpokenElyza to facilitate future research in Japanese spoken dialog evaluation\footnote{\url{https://huggingface.co/datasets/sbintuitions/voicebench-ja}}.

\begin{figure}[t]
    \centering
    \includegraphics[width=0.48\textwidth]{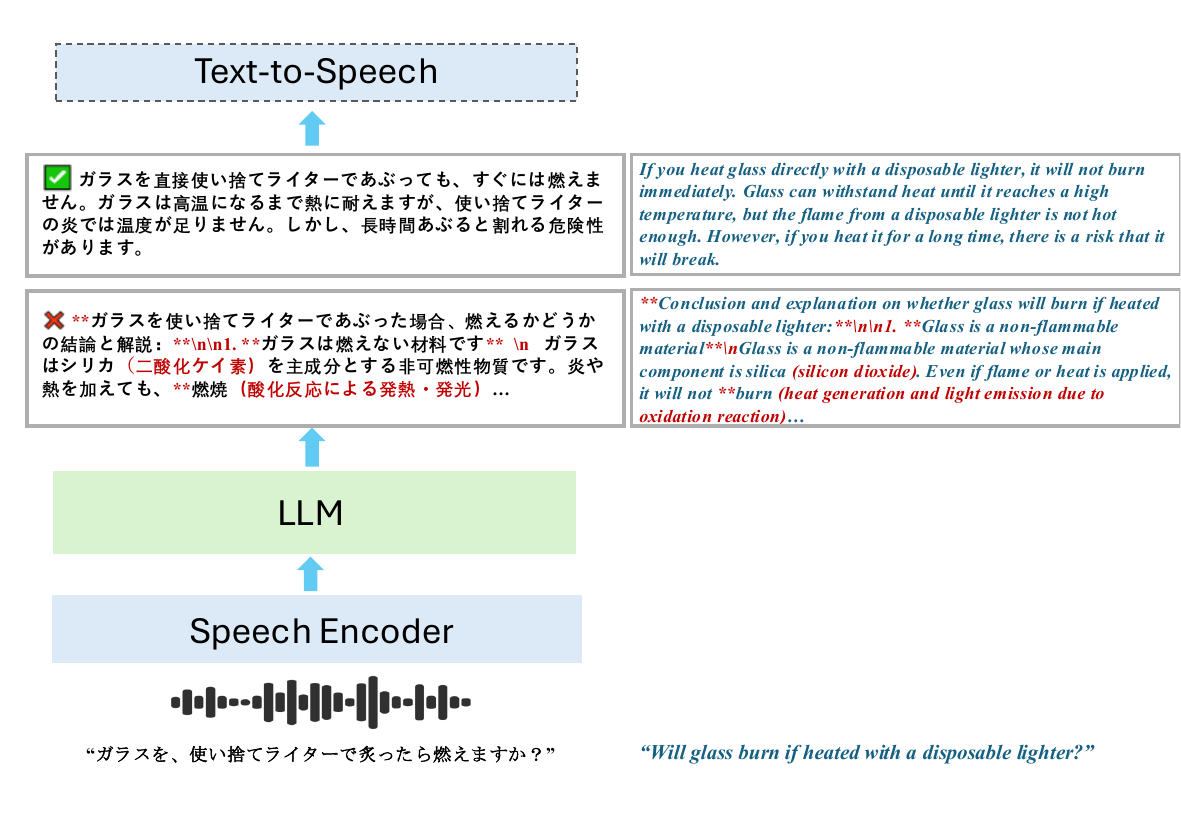}
    \caption{Overview of our speech-worthy alignment. SpeechLLM responses are aligned from written style (verbose, symbol-heavy) to spoken style (concise, conversational) for auditory comprehension and TTS synthesis. English translations at right; non-speech-worthy elements are highlighted in red.}
    \label{fig:my_image}
\end{figure}

\section{Method}
\subsection{Model Architecture}
Our model follows a widely-used SpeechLLM architecture similar to Qwen2-Audio \cite{Qwen2-Audio}, where a shallow projector layer connects a speech encoder (from Whisper-large \cite{radford2023robust}) with an LLM. The LLM consumes projected representations, optionally augmented with text embeddings (e.g., of system prompts), to generate outputs. We use Sarashina-7B\footnote{We use our internal weights; similar models are at \url{https://huggingface.co/sbintuitions/sarashina2-7b}} as our base LLM to utilize its strong Japanese language capabilities.

\subsection{Preference-Based Alignment}
\label{sec:alignment}
Our goal is to adapt SpeechLLMs to produce speech-worthy outputs while preserving their instruction-following capabilities. We frame this as a preference optimization problem, where speech-worthy responses are preferred over written-style responses.

\textbf{Direct Preference Optimization (DPO).} We employ DPO \cite{rafailov2023direct} to directly optimize a policy to satisfy the preference on speech-worthy without requiring a separate reward model. Given preference pairs $(y_w, y_l)$ where $y_w$ is the speech-worthy (preferred) response and $y_l$ is the written-style (dispreferred) response, DPO minimizes:
{\footnotesize\begin{equation}
\begin{split}
    \mathcal{L}_{\text{DPO}}(\pi_\theta; \pi_{\text{ref}}) &= -\mathbb{E}_{(x, y_w, y_l) \sim \mathcal{D}} \bigg[ \\
    &\quad \log \sigma \left( \beta \log \frac{\pi_\theta(y_w \mid x)}{\pi_{\text{ref}}(y_w \mid x)} - \beta \log \frac{\pi_\theta(y_l \mid x)}{\pi_{\text{ref}}(y_l \mid x)} \right) \bigg]
\end{split}
\end{equation}
}
where $\pi_\theta$ is the policy being optimized, $\pi_{\text{ref}}$ is the reference model (our pretrained checkpoint), $\sigma$ is the logistic sigmoid function, and $\beta$ controls the strength of the KL divergence constraint. Intuitively, DPO increases the relative likelihood of preferred responses while staying close to the reference distribution to prevent degeneration, in our case, deterioration on instruction following ability.

\textbf{Combined DPO + SFT Training.} We combine DPO with supervised fine-tuning (SFT) on chosen responses, and the SFT loss directly maximizes the likelihood of speech-worthy responses:
\begin{equation}
    \mathcal{L}_{\text{SFT}}(\pi_\theta) = -\mathbb{E}_{(x, y_w) \sim \mathcal{D}} \left[ \log \pi_\theta(y_w\mid x) \right]
\end{equation}
The combined objective is $\mathcal{L} = w \cdot \mathcal{L}_{\text{DPO}} + (1-w) \cdot \mathcal{L}_{\text{SFT}}$, where $w$ controls the relative importance of preference learning versus supervised learning. We experiment with different $w$ values in our experiments (cf. Section \ref{sec:settings}).

\subsection{Constructing SpokenElyza}
\label{sec:spokenelyza}
To the best of our knowledge, there are no existing benchmarks to rigorously evaluate the speech-worthiness of Japanese SpeechLLMs. We introduce \textbf{SpokenElyza}, a high-quality evaluation benchmark derived from the standard  ELYZA-tasks-100 dataset \cite{elyzatasks100}. While  ELYZA-tasks-100 is widely used for assessing Japanese instruction-following capabilities, it targets text LLMs and contains several examples unsuitable for spoken interaction (e.g., text string manipulation). 
We constructed SpokenElyza through a four-stage curation pipeline designed to ensure both linguistic naturalness and acoustic suitability:

\textbf{Modality filtering}. We screened the original 100 examples to filter out tasks that are inherently non-verbalizable (e.g., text string processing, email/essay writing). We also refined input instructions to reflect natural spoken queries. This filtering process yielded a core set of 36 instruction-response pairs, which we call \textbf{Elyza} (the filtered subset with original written-style responses). Based on Elyza, we then apply speech-worthy style transfer to create \textbf{SpokenElyza} via the next three steps.

\textbf{Speech-Worthy Style Transfer}. The original ground-truth responses in Elyza exhibit strong written-text biases, including markdown formatting, bullet points, numbered lists, and excessive length. We employed a strong LLM (gpt-oss-120B) to rewrite responses under strict speech-friendly constraints: (1) removal of all markdown formatting and lists, (2) conversion of written predicates to polite spoken forms, (3) simplification of complex nested sentences, and (4) ensuring comprehensibility \emph{solely through audio without visual aids}. Two examples required responses that remained excessively long even after rewriting due to the inherent complexity of the underlying tasks; we excluded these from SpokenElyza, resulting in 34 examples. We retain all 36 examples in Elyza, as the written-style evaluation does not penalize length.

\textbf{Auditory Verification by Native Experts.} To ensure SpokenElyza is valid for \emph{listening} rather than just reading, we introduced a human-in-the-loop verification step. We synthesized the rewritten responses using our in-house TTS system. A native Japanese researcher listened to each audio sample and made corrections if needed, ensuring auditory intelligibility.

\textbf{Evaluation Criteria Standardization}. Individual Elyza examples contain example-specific evaluation aspects that sometimes directly assign absolute scores, potentially conflicting with LLM-as-judge prompts. We converted these absolute scores into relative deductions from the maximum score of 5, ensuring consistent evaluation across all the examples.

We release SpokenElyza as a benchmark to facilitate future research in Japanese spoken dialog evaluation.

\section{Experiments}
\textbf{Pretraining Datasets.}
Our first step is pretraining to align the speech and text modalities. We freeze both the LLM and the audio encoder, training only the projector layer. For this alignment phase, we use the large-scale open-source ReazonSpeech dataset \cite{reazonspeech2023}, which provides Japanese speech-text pairs. To further scale up pretraining data, we also incorporate our in-house datasets containing paired (audio, text) examples.

\textbf{Preference Training Datasets}. After aligning the auditory and text modalities into a shared representation space, we adapt the model's written-style outputs to be suitable for spoken conversation via DPO. We next introduce the preference training datasets:
\begin{itemize}
\item{} \textbf{SpeechPref} \cite{cho-etal-2024-speechworthy}. SpeechPref is a large-scale preference corpus specifically curated for speech. Unlike text-oriented instruction-tuning datasets like Alpaca \cite{alpaca}, SpeechPref is constructed to be ``audio-first'' \cite{cho-etal-2024-speechworthy}. It is derived from Databricks-Dolly-15K \cite{DatabricksBlog2023DollyV2}, filtered to exclude examples unsuitable for speech-based interactions such as coding and mathematical derivations. Annotators explicitly listened to TTS-synthesized waveforms of different responses and assigned preferences, ensuring that auditory-specific nuances like rhythm are considered and non-verbalizable formatting is excluded. We translate SpeechPref to Japanese and synthesize instruction waveforms  using our in-house TTS model.

\item{} \textbf{InstructS2S-200K} \cite{fang2024llama}. We further expand our preference training data by adding InstructS2S-200K, originally designed for training unified speech-to-speech systems like Llama-Omni \cite{fang2024llama}. We translated the instructions into Japanese to obtain a diverse array of spoken instructions across various intents and domains. For each instruction, we generate multiple response rollouts using the pretrained checkpoint and employ gpt-oss-120B  to assign speech-suitability scores ($S \in [0, 100]$). To ensure high-quality preference pairs for DPO, we apply a rigorous margin-based filtering protocol: we discard instructions where the maximum rollout score is below 90, use the highest-scoring rollout as the chosen response ($y_w$), and retain a rollout as rejected ($y_l$) only if it satisfies $S_{\text{rejected}} \times 1.5 < S_{\text{chosen}}$. This ensures the DPO objective is optimized exclusively on pairs with large gaps in speech-worthiness.
Then we synthesize instruction waveforms using our in-house TTS model.

\item{} \textbf{DeepDialog} \cite{deepdialog}. Both SpeechPref and InstructS2S-200K are instruction-following datasets. We further incorporate DeepDialog, which contains daily chit-chat spoken conversations. We translate the first turn of each dialog and apply the same rollout-based preference construction: the highest-scoring rollout becomes the chosen response, while lower-scoring rollouts serve as rejected examples, followed by waveform synthesis.
\end{itemize}

\textbf{Evaluation Datasets.}
We evaluate on two benchmarks: (1) \textbf{Elyza}, using the written-style responses and evaluation rubric from  ELYZA-tasks-100 \cite{elyzatasks100}; and (2) \textbf{SpokenElyza}, using our speech-worthy responses and the rubric in Table~\ref{tab:scoring_rubric}. Comparing performance on both allows us to assess the trade-off between speech-worthy alignment and general instruction-following capability.

\subsection{Evaluation Metrics}

\begin{table}[t]
    \centering
    \scriptsize
    \caption{
    Scoring rubric used for the LLM-as-Judge evaluation in SpokenElyza.
    The response to each instruction is scored from 1 to 5; we report averaged score among the examples.
    }
    \label{tab:scoring_rubric}
    \begin{tabular}{c p{0.82\linewidth}}
        \toprule
        \textbf{Score} & \textbf{Description} \\
        \midrule
        \textbf{1} & \textbf{Fatal Failure.} The response contains fundamental factual errors, hallucinations, or completely fails to follow the instruction. \\
        \midrule
        \textbf{2} & \textbf{Low Quality.} The response contains partial errors, incoherent Japanese, or excessive safety refusals (e.g., declining to answer harmless queries). \\
        \midrule
        \textbf{3} & \textbf{Written Style.} The content is factually accurate, but the format is unsuitable for speech, containing markdown, bullet points, URLs, or visual references (e.g., "as shown below"). \\
        \midrule
        \textbf{4} & \textbf{Spoken Style.} The response is accurate and composed in natural spoken Japanese. It is free of non-verbalizable artifacts and structurally ready for synthesis. \\
        \midrule
        \textbf{5} & \textbf{Auditory Ideal.} In addition to meeting the Score 4 criteria, the response is stellar in its listenability. It uses short sentences to reduce cognitive load and incorporates natural conversational markers to enhance the tone. \\
        \bottomrule
    \end{tabular}
\end{table}

\textbf{LLM-as-Judge.} We conduct evaluations via LLM-as-judge \cite{zheng2023judging} using Qwen2.5-32B-Instruct, which aligns closely with human judgment for open-ended generation tasks \cite{liu-etal-2023-g}. For Elyza, we use its standard evaluation prompt focused on factual accuracy and instruction following. For SpokenElyza, we define a speech-specific scoring rubric (Table~\ref{tab:scoring_rubric}) that evaluates whether responses are suitable for auditory consumption: 
factually correct content presented in written style (e.g.,  markdown, URLs) still receives lower scores as it cannot be naturally verbalized by TTS systems.

\textbf{Automatic Surface-Form Metrics.}
Following \cite{cho-etal-2024-speechworthy}, we further conduct  surface-form evaluations to assess the speech-suitability of generated responses. They serve as proxies for three aspects of spoken dialog:

\textbf{Conciseness (Word Count)}. Spoken information processing is serial and transient, and concise responses are generally preferred \cite{cho-etal-2024-speechworthy}. We calculate total word count by tokenizing Japanese sentences using the Janome tokenizer.\footnote{\texttt{https://github.com/mocobeta/janome}} Lower word counts correlate with higher listenability \cite{cho-etal-2024-speechworthy}.
\textbf{Sentence Complexity (Dependency Depth)}. We measure syntactic complexity using dependency depth. Following \cite{cho-etal-2024-speechworthy}, we use SpaCy's Japanese dependency parser\footnote{\texttt{https://spacy.io/models/ja}. We report the averaged performance using \texttt{ja\_core\_news\_sm}, \texttt{ja\_core\_news\_md}, \texttt{ja\_core\_news\_lg}.} to construct the dependency graph of each sentence and calculate the maximum depth. Lower depth indicates simpler sentence structures that are easier to comprehend when listening.
\textbf{Non-Vocalizable Content (NV\%).} LLMs often output artifacts unsuitable for TTS systems. We quantify this by measuring the percentage of non-vocalizable characters. Lower NV\% indicates responses better suited to the speech modality.

\subsection{Experimental Settings}
\label{sec:settings}
For the \textbf{pretraining} stage, we use a learning rate of 1e-4 and pretrain for 100K steps using 32 H100 GPUs with per-GPU batch size of 16. This phase only trains the projector layer.

For the \textbf{preference training} stage, we investigate three axes of variation. Firstly, we vary the \textbf{DPO loss weight} $w \in \{0.5, 0.9, 0.95, 0.99\}$, controlling the balance between preference learning and SFT. Next, we vary the \textbf{trainable parameters} in addition to the projector layer, comparing two strategies:
\textbf{TopLayers}: Fine-tuning the top four transformer layers of the LLM, a common heuristic for adaptation tasks. 
\textbf{KQ-LN}: Fine-tuning key, query, and LayerNorm parameters across all transformer layers. Our hypothesis is that style transfer (written $\rightarrow$ spoken) primarily requires adjusting attention patterns and output distributions rather than learning new knowledge.
Lastly, following \cite{cho-etal-2024-speechworthy}, we also employ a \textbf{spoken system prompt} to explicitly instruct the model to generate speech-friendly conversational responses. We show in our experiments that the spoken system prompt yields synergistic improvements.
We use a learning rate of 5e-6 and train for 2 epochs. Each preference training  run uses 8 H100 GPUs.

\section{Results}
\subsection{Main Results}
\begin{table}[t]
\centering\scriptsize
\caption{LLM-as-Judge scores on Elyza and SpokenElyza. PT: Pretrained model. Prpt: Spoken system prompt. Higher is better; maximum score is 5. Best results per benchmark in \textbf{bold}.}
\renewcommand{\arraystretch}{1.3}
\begin{tabular}{r|c|c}
\hline
 & Elyza & SpokenElyza \\
\hline
PT & \textbf{3.97} & 2.91 \\
\hline
PT + Prpt & - & 2.94 \\
\hline
DPO + SFT & 3.78 & 2.97 \\
\hline
DPO + SFT + Prpt & - & \textbf{3.44} \\
\hline
\end{tabular}
\label{tab:llmjudgeres}
\end{table}

\textbf{LLM-as-Judge Results.} Table~\ref{tab:llmjudgeres} demonstrates the effectiveness of speech-worthy alignment for SpeechLLMs. On Elyza, preference-trained models show a modest performance decrease compared to the pretrained checkpoint (3.78 vs. 3.97, a 5\% relative drop),  which is expected for two reasons: (1) prior work has shown that SpeechLLMs exhibit reduced capabilities compared to cascaded systems (ASR+LLM+TTS) \cite{chen2024voicebench}, and (2) Elyza is designed for text-based LLMs and does not distinguish between written and spoken styles. 

On SpokenElyza, we observe that adding the spoken system prompt slightly helps the pretrained model (2.91 $\rightarrow$ 2.94). Combining DPO+SFT training with the spoken prompt achieves the best performance (3.44), representing an 18\% relative improvement over the pretrained baseline. This demonstrates the complementary benefits of preference training and prompting, consistent with findings for text LLMs \cite{woo-etal-2025-think}.

\begin{table}[t]
\centering\scriptsize
\renewcommand{\arraystretch}{1.3}
\caption{Surface-form evaluation results on SpokenElyza. PT:  Pretrained model, Prpt: spoken system prompt. DD: dependency depth. NV\%: percentage of non-verbalizable contents.}
\begin{tabular}{rrrrr}
\hline
  & \textbf{Model} & \textbf{Word Count $\downarrow$} & \textbf{DD $\downarrow$} & \textbf{NV \% $\downarrow$} \\ \hline
\multirow{4}{*} & PT & 325.91 & 6.38 & 13.46\% \\
 & DPO + SFT & 302.15 & 6.42 & 12.65\% \\
 & PT + Prpt & \textbf{65.53} & 5.06 & 3.69\% \\
 & DPO + SFT + Prpt & 77.79 & \textbf{4.97} & \textbf{3.24\%} \\ \hline
\end{tabular}
\label{tab:surfaceres}
\end{table}

\textbf{Surface-Form Results} are shown in Table~\ref{tab:surfaceres}. The pretrained checkpoint generates lengthy responses averaging 326 words, with high dependency depth (6.38) and substantial non-vocalizable content (13.46\%). These characteristics reflect the lengthy and complex responses from common text LLMs, which are problematic for TTS synthesis and auditory comprehension.
Adding the spoken system prompt dramatically reduces verbosity (326 $\rightarrow$ 66 words) and improves both dependency depth and NV\%. The preference-trained model with prompting achieves the smallest dependency depth (4.97) and lowest NV\% (3.24\%), though with slightly higher word count than prompting alone (78 vs. 66). This suggests that preference training encourages more complete responses while maintaining speech-appropriate structure, whereas prompting alone may overly truncate outputs.

\subsection{Ablation Studies}
\begin{table}[t]
\centering\scriptsize
\caption{Comparison of parameter selection strategies. KQ-LN (key-query and LayerNorm) outperforms TopLayers tuning on both benchmarks.}
\renewcommand{\arraystretch}{1.2}
\begin{tabular}{r|c|c}
\hline
 & Elyza & SpokenElyza \\
\hline
Pretrained & \textbf{3.97} & 2.91 \\
\hline\midrule
TopLayers & 3.61 & 2.91 \\
\hline
TopLayers + Prpt & - & 3.38 \\
\hline
KQ-LN & \underline{3.78} & 2.97 \\
\hline
KQ-LN + Prpt & - & \textbf{3.44} \\
\hline
\end{tabular}
\label{tab:trainconfig}
\end{table}

\textbf{Parameter Selection Strategy.} Table~\ref{tab:trainconfig} compares our two parameter selection strategies. In addition to the projector layer, we explore tuning the top four LLM transformer layers (TopLayers) or tuning key-query and LayerNorm parameters (KQ-LN), as discussed in Section \ref{sec:settings}.
We observe that KQ-LN outperforms top-layer tuning on both benchmarks. On Elyza, KQ-LN shows a smaller performance drop compared to the pretrained baseline (3.78 vs. 3.97, or 5\% relative) than TopLayers (3.61, or 9\% relative). On SpokenElyza with prompting, KQ-LN achieves 3.44 vs. 3.38 for TopLayers.
These results support our hypothesis that style transfer from written to spoken primarily requires adjusting the transformer layers' attention patterns and output distributions, rather than modifying the encoded factual knowledge. KQ-LN modifies how the model attends to and normalizes representations across all layers, altering the style while preserving learned knowledge.

\begin{figure}[t]
    \centering
    \includegraphics[width=0.42\textwidth]{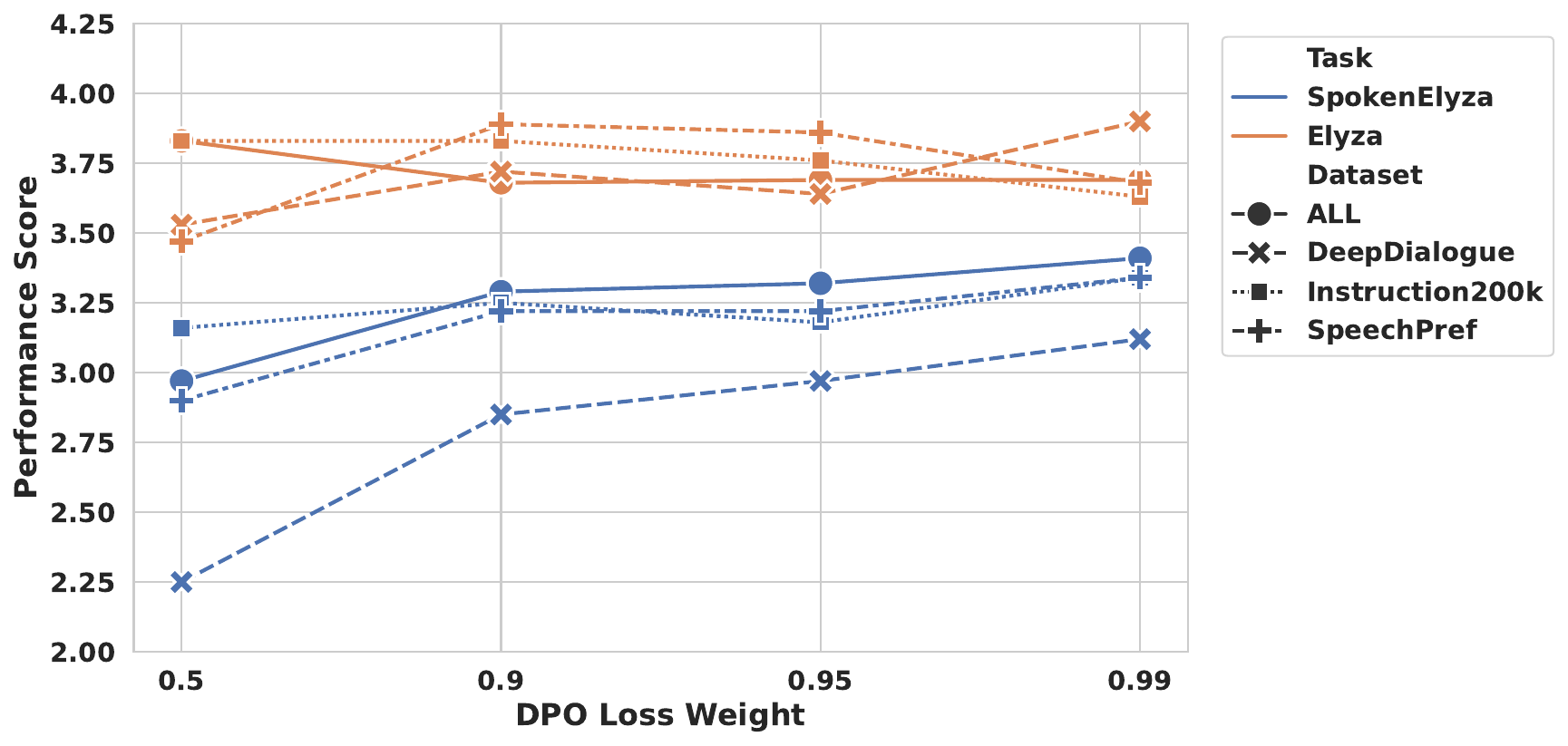}
    \caption{Effect of DPO loss weight on LLM-as-Judge scores. Higher DPO weights improve SpokenElyza performance across all datasets. ALL: combination of all preference training datasets.}
    \label{fig:varyweights}
\end{figure}

\textbf{Effect of DPO Loss Weight.} Figure~\ref{fig:varyweights} shows performance across DPO loss weights $w \in \{0.5, 0.9, 0.95, 0.99\}$ for individual datasets and their combination. On SpokenElyza, increasing the DPO weight shows consistent improved performance across all datasets (except for InstructS2S-200K when $w=0.95$). This effect is most pronounced for DeepDialog, a casual chit-chat dataset where intensive SFT causes the model to lose instruction-following abilities. These results suggest that learning the \emph{contrast} between spoken and written styles via DPO is more important than simply imitating speech-worthy responses via SFT.

On Elyza, the trends are less consistent. For the combined dataset (ALL), performance decreases slightly from $w=0.5$ to $w=0.9$ and then plateaus. From $w=0.9$ to $w=0.99$, we observe a decreasing trend when using InstructS2S-200K and SpeechPref, indicating that higher DPO weights push the model further toward spoken style, which Elyza's text-oriented evaluation penalizes. Balancing improved speech-worthiness against degraded
text-benchmark performance requires careful tuning.

\section{Conclusion}
We proposed a preference-based alignment approach to adapt Japanese SpeechLLMs for speech-worthy outputs. 
To rigorously evaluate this task, we introduced SpokenElyza, the first benchmark for Japanese speech-worthiness, constructed through modality filtering, style transfer, and auditory verification by native experts. We will release SpokenElyza to support future research on Japanese spoken dialog systems.

Our experiments show that applying DPO combined with SFT achieves substantial improvement on SpokenElyza while largely preserving performance on the text-oriented Elyza benchmark. 
We also observed that higher DPO loss weights improve speech-worthiness, and that tuning key-query and LayerNorm parameters is more effective than top-layer tuning for this style transfer task.

Our work focuses on the Japanese language, which has particularly pronounced written-spoken divergence; future work may explore
generalizations to other languages.

\section{Generative AI Use Disclosure}
Generative AI tools (e.g., Gemini) were used for language editing and improving the phrasing of this manuscript.

\bibliographystyle{IEEEtran}
\bibliography{mybib}

\end{document}